\documentclass[aps,showpacs,nofootinbib,amsmath,amssymb,preprint]{revtex4}
\usepackage{bm}
\usepackage{amsthm}

\begin{document}

\title{
Note on the parametrized black hole quasinormal ringdown formalism
}

\author{
Masashi Kimura${}^{1}$
}

\affiliation{
${}^{1}$Department of Physics, Rikkyo University, Tokyo 171-8501, Japan
}

\date{\today}
\preprint{RUP-20-2}

\begin{abstract}
The parametrized black hole quasinormal ringdown formalism is useful
to compute quasinormal mode (QNM) frequencies if a master equation for the gravitational perturbation 
around a black hole
has a small deviation from the Regge-Wheeler or Zerilli equation.
In this formalism, the deviation of QNM frequency from general relativity can be calculated by
small deviation parameters and model independent coefficients.
In this paper, we derive recursion relations for the model independent coefficients.
Using these relations, the higher order coefficients are written only by the lower order coefficients.
Thus, we only need the lower order coefficients when 
we numerically compute the model independent coefficients.
\end{abstract}

\maketitle

\section{Introduction}
The master equations for the gravitational perturbation 
around the Schwarzschild black holes are known as
the Regge-Wheeler or Zerilli equation~\cite{Regge:1957td, Zerilli:1970se}.
If we consider a gravity theory whose deviation from general relativity (GR) is very small,
the master equations for the gravitational perturbation around
spherically symmetric black holes sometimes
have only small deviations from the Regge-Wheeler or Zerilli equation in the form:
\begin{align}
f \frac{d}{dr} \left(f\frac{d}{dr} \Phi \right)+ (\omega^2 - f (V^{(0)} + \delta V) )\Phi = 0,
\label{mastereq}
\end{align}
with 
\begin{align}
f &= 1 - \frac{r_H}{r}.
\end{align}
The back ground effective potential $V^{(0)}$ is
\begin{align}
V^{(0)} &= V_+  = \frac{9 \lambda r_H^2 r + 3 \lambda^2 r_H r^2 + \lambda^2 (\lambda +2)r^3 + 9 r_H^3}{r^3(\lambda r + 3 r_H)^2},
\label{evenv}
\end{align}
where $\lambda = \ell^2 + \ell -2$, for the even parity perturbation, and
\begin{align}
V^{(0)} &= V_-  = \frac{\ell (\ell + 1)}{r^2} - \frac{3 r_H}{r^3},
\label{oddv}
\end{align}
for the odd parity perturbation.
$\delta V$ denotes the small deviation of the effective potential from the GR case.
In~\cite{Cardoso:2019mqo}, the form of $\delta V$ is assumed to be
\begin{align}
\delta V = \delta V_\pm &= \frac{1}{r_H^2}\sum_{j = 0}^\infty \alpha_j^\pm \left(\frac{r_H}{r}\right)^j,
\label{deltav}
\end{align}
where $\alpha_j^\pm$ are small parameters which depend on the details of gravity theories
or physical situations.
Note that many models are included in this parameterization, {\it e.g.,}~\cite{Cardoso:2019mqo, Cardoso:2018ptl, McManus:2019ulj, Tattersall:2019nmh}.
The parametrized black hole quasinormal ringdown formalism~\cite{Cardoso:2019mqo} 
is useful to compute quasinormal mode (QNM) frequencies of the above systems Eqs.~\eqref{mastereq}-\eqref{deltav}.
At the first order of the small parameters $\alpha_j^\pm$, the QNM frequency becomes 
\begin{align}
\omega_{\rm QNM} &= \omega_0 + \delta \omega 
\notag\\&= \omega_0^\pm + \sum_{j = 0}^\infty \alpha_j^\pm e_j^\pm,
\label{qnmomegaformula}
\end{align}
where $\omega_0 = \omega_0^\pm$
are the QNM frequencies for the GR case, {\it i.e.,} for the Schwarzschild black holes,
and $\delta \omega = \sum_{j = 0}^\infty \alpha_j^\pm e_j^\pm$ denote the 
deviation from the GR case.
The coefficients $e_j^\pm$ are constants which do not depend on $\alpha_j^\pm$.
In this sence, $e_j^\pm$ can be considered as the model independent coefficients.
In the previous work~\cite{Cardoso:2019mqo}, the coefficients $e_j^\pm$ were calculated numerically, but 
to obtain them in a high accuracy was technically not easy,
especially for large $j$.
In this paper, we derive recursion relations among
$e_j^\pm$ with different $j$.
Using these relations, we can write $e_j^\pm$ only from low $j$ coefficients.

\section{ambiguity of the effective potential at the first order of small parameters}
\label{secii}
Introducing a new master variable $\tilde{\Phi}$ as
\begin{align}
\Phi = (1  + \epsilon X) \tilde{\Phi} + \epsilon Y f \frac{d\tilde{\Phi}}{dr}
\end{align}
where $\epsilon$ is a small parameter, which is assumed to be same order as $\alpha_j^\pm$, 
and $X, Y$ are functions of $r$.
If we choose $X$ as
\begin{align}
X = c_1 - \frac{f}{2} \frac{dY}{dr},
\end{align}
with a constant $c_1$,
Eq.~\eqref{mastereq} becomes
\begin{align}
f \frac{d}{dr} \left(f\frac{d}{dr} \tilde{\Phi} \right)+ (\omega^2 - f (V^{(0)} + \delta V+ \delta W) ) \tilde{\Phi} = 0,
\label{mastereq2}
\end{align}
with
\begin{align}
\delta W &= 
\epsilon\left[
- Y\frac{d(f V^{(0)})}{dr}
+
2 (\omega_0^2 - fV^{(0)})\frac{dY}{dr}
+
\frac{1}{2}
\frac{d}{dr} \left(f \frac{d}{dr} \left(f \frac{d Y}{dr}\right)\right)
\right],
\label{deltaw}
\end{align}
at the first order of $\epsilon$.
In the above derivation, 
we used the relation $\tilde{\Phi} = \Phi  + {\cal O}(\epsilon)$
and that $\Phi$ satisfies Eq.~\eqref{mastereq}.
Unless the choice of $X$ and $Y$ do not change the QNM boundary conditions between
$\Phi$ and $\tilde{\Phi}$,
$\delta W$ can be considered as an ambiguity of the effective potential.
(see also Appendix.~\ref{generalcase} for general discussion of the ambiguity of the effective potential for 
master equations with Schr\"odinger form.)

\section{recursion relation for the model independent coefficients}
In this section, we derive recursion relations among $e_j^\pm$ with different $j$.
Let us consider the GR case, {\it i.e.,} $\delta V = 0$, where the master equation does not contain any correction term,
\begin{align}
f \frac{d}{dr} \left(f\frac{d}{dr} \Phi \right) + (\omega^2 - f V^{(0)} ) \Phi = 0.
\label{zerothmastereq}
\end{align}
From the discussion in the previous section, this equation can be rewritten in the form of Eq.~\eqref{mastereq2} with 
$\delta V = 0$.

\subsection{Odd parity case}
For the odd parity case, the back ground effective potential 
$V^{(0)}$ becomes $V_-$ in Eq.~\eqref{oddv}.
Setting the function $Y$ as
\begin{align}
Y = y_j \left(\frac{r_H}{r}\right)^j,
\end{align}
with an integer $j$ and a constant $y_j$,
$\delta W$ in Eq.~\eqref{deltaw} becomes
\begin{align}
\delta W &=
\epsilon y_j  \bigg[- \frac{2 j (\omega_0^-)^2}{r_H} \left(\frac{r_H}{r}\right)^{j+1}
-
\frac{(1+j)(-4\ell (\ell +1) + j (j+2))}{2 r_H^3}
\left(\frac{r_H}{r}\right)^{j+3}
\notag\\& \quad +
\frac{(2 j + 3)(-6 - 2 \ell (\ell +1) + j (j + 3))}{2 r_H^3}
\left(\frac{r_H}{r}\right)^{j+4}
-
\frac{(j-2)(j+2)(j+6)}{2 r_H^3}
\left(\frac{r_H}{r}\right)^{j+5}
\bigg].
\end{align}
We can read $\alpha_j^-$ as
\begin{align}
\alpha_{j+1}^- &= - \frac{2 j (\omega_0^-)^2}{r_H}\epsilon y_j,
\\
\alpha_{j+3}^- &=
-
\frac{(1+j)(-4\ell (\ell +1) + j (j+2))}{2 r_H^3} \epsilon y_j,
\\
\alpha_{j+4}^- &=
\frac{(2 j + 3)(-6 - 2 \ell (\ell +1) + j (j + 3))}{2 r_H^3} \epsilon y_j,
\\
\alpha_{j+5}^- &=
-
\frac{(j-2)(j+2)(j+6)}{2 r_H^3} \epsilon y_j,
\end{align}
and the QNM frequency at the first order becomes
\begin{align}
\omega_{\rm QNM} = \omega_0^- + 
\alpha_{j+1}^- e_{j+1}^-
+
\alpha_{j+1}^- e_{j+3}^-
+
\alpha_{j+1}^- e_{j+4}^-
+
\alpha_{j+1}^- e_{j+5}^-.
\label{qnm1}
\end{align}
However, because now we are studying the system with Eq.~\eqref{zerothmastereq}, the QNM frequency Eq.~\eqref{qnm1} should be same as the GR case, {\it i.e.,} $\omega_{\rm QNM}= \omega_0^-$.
Thus, the
recursion relation
\begin{align}
\alpha_{j+1}^- e_{j+1}^-
+
\alpha_{j+1}^- e_{j+3}^-
+
\alpha_{j+1}^- e_{j+4}^-
+
\alpha_{j+1}^- e_{j+5}^- = 0
\label{receq}
\end{align}
should hold.
Eq.~\eqref{receq} can be explicitly written as
\begin{align}
& - 4 j r_H^2 (\omega_0^-)^2 e_{j+1}^- 
-
(j+1)(-4\ell (\ell +1) + j (j+2)) e_{j+3}^-
\notag\\& +
(2 j + 3)(-6 - 2 \ell (\ell +1) + j (j + 3))e_{j+4}^-
-
(j-2)(j+2)(j+6) e_{j+5}^- = 0.
\label{receq2}
\end{align}
We should note that while $Y \sim r$ at $r \to \infty$ for $j = -1$, 
the QNM boundary condition for $\tilde{\Phi}$ still corresponds 
to the QNM boundary condition for the original master variable $\Phi$.
Thus, we can put $j \ge -1$ to the above recursion relation,
and we conclude that all $e_j^-$ with $j \ge 3$ can be written by
$e_0^-, e_2^-$ and $e_7^-$.\footnote{\label{foot1}
Note that $e_7^-$ cannot be written by $e_0^-, e_2^-$ because 
the coefficient of $e_7^-$ in Eq.~\eqref{receq2} with $j = 2$ vanishes.
}
For example, the relations
\begin{align}
e_3^- &=
\frac{2 (\ell -1)\ell (\ell + 1)(\ell +2)(\ell^2 + \ell - 1) r_H^2 (\omega_0^-)^2 }{(\ell -1)^2 \ell^2 (\ell + 1)^2 (\ell + 2)^2 + 36 r_H^2 (\omega_0^-)^2} e_0^-
- 
\frac{12(\ell^2 + \ell -2)r_H^2 (\omega_0^-)^2 }{(\ell -1)^2 \ell^2 (\ell + 1)^2 (\ell + 2)^2 + 36 r_H^2 (\omega_0^-)^2}
e_2^-,
\label{e3m}
\\
e_4^- &= - \frac{4}{15} r_H^2 (\omega_0^-)^2 e_0^-
+
\frac{2}{15}(4 + \ell + \ell^2) e_3^-,
\\
e_5^- &= - \frac{1}{15} (3 + \ell + \ell^2)r_H^2 (\omega_0^-)^2 e_0^-
+
\frac{1}{30}(12 + \ell(\ell + 1)(2 + \ell + \ell^2)) e_3^-,
\\
e_6^- &= 
-\frac{2}{35}(3+\ell + \ell^2)r_H^2 (\omega_0^-)^2 e_0^-
+
\frac{\ell(\ell+1)(\ell^2(\ell+1)^2 + 8) - 4(6+5 r_H^2 (\omega_0^-)^2)}{35(\ell^2 + \ell - 2)}e_3^-.
\\
\cdots & 
\notag
\end{align}
hold.\footnote{
One may wonder how to express $e_3^-$ by $e_0^-$ and $e_2^-$.
Eq.~\eqref{receq2} with $j = -1,0,1,2$ give four equations which contain $e_0^-, e_2^-, e_3^-, e_4^-, e_5^-, e_6^-$.
Solving them with respect to $e_3^-, e_4^-, e_5^-, e_6^-$,
we can write $e_3^-$ by $e_0^-$ and $e_2^-$.
}
We report that $e_j^-$ with $j\ge 3$ evaluated from the numerical data of $e_0^-, e_2^-, e_7^-$ in~\cite{Cardoso:2019mqo} agree with that in the previous work.

\subsection{Even parity case}
For the even parity case, the back ground effective potential 
$V^{(0)}$ becomes $V_+$ in Eq.~\eqref{evenv}.
Setting the function $Y$ as
\begin{align}
Y = y_j \frac{(\lambda r + 3 r_H)^3}{r_H^3}\left(\frac{r_H}{r}\right)^j,
\end{align}
with an integer $j$ and a constant $y_j$,
$\delta W$ becomes
\begin{align}
\delta W &=
\frac{\epsilon y_j}{r_H^3}  \bigg[
-2 \lambda ^3 (j-3) r_H^2
   (\omega_0^+)^2
\left(\frac{r_H}{r}\right)^{j-2}
-18 \lambda ^2 (j-2) r_H^2 (\omega_0^+)^2
\left(\frac{r_H}{r}\right)^{j-1}
\notag \\ & \quad
+
\left(
\frac{1}{2} \lambda ^3 (j-2)
   (4 \lambda -(j-4) j+5)
-54
   \lambda  (j-1) r_H^2 (\omega_0^+)^2
\right)
\left(\frac{r_H}{r}\right)^{j}
\notag \\ & \quad
+
\left(
\frac{1}{2} \lambda ^2  \left(6
   (\lambda -1) \lambda +(2 \lambda -9) j^3-9 (\lambda -3)
   j^2+((25-4 \lambda ) \lambda +6) j\right)-54 j
   r_H^2 (\omega_0^+)^2
\right)
\left(\frac{r_H}{r}\right)^{j+1}
\notag \\ & \quad
+\frac{\lambda 
   \left(3 (\lambda -4) \lambda -((\lambda -18)
   \lambda +27) j^3+3 (\lambda -9) \lambda  j^2+((21-23
   \lambda ) \lambda +27) j\right)}{2
 }
\left(\frac{r_H}{r}\right)^{j+2}
\notag \\ & \quad
\frac{9 \left(\lambda ^2 \left(-\left(j^3+4
   j+2\right)\right)+\lambda  (j+1) \left(6 j^2+3
   j+4\right)-3 n (j+1) (j+2)\right)}{2 }
\left(\frac{r_H}{r}\right)^{j+3}
\notag \\ & \quad
+\frac{9 (3 (j+1) (j+2) (2 j+3)-\lambda 
   (j (3 j (j+3)+13)+9))}{2 }
\left(\frac{r_H}{r}\right)^{j+4}
\notag \\ & \quad
-\frac{27 (j+2)^3 }{2
   }
\left(\frac{r_H}{r}\right)^{j+5}
\bigg].
\end{align}
We can read $\alpha_{j-2}, \alpha_{j-1}, \cdots, \alpha_{j+5}$
by comparing with
$\delta W = r_H^{-2}\sum_{j = 0}^\infty \alpha_j^\pm (r_H/r)^j$, 
and the QNM frequency becomes
\begin{align}
\omega_{\rm QNM} = \omega_0^+ + 
\sum_{k = j-2}^{j+5}
\alpha_{k}^+ e_{k}^+.
\label{qnm1even}
\end{align}
Similar to the odd parity case, 
because 
the QNM frequency \eqref{qnm1even} should be same as the GR case, the
recursion relation
\begin{align}
\sum_{k = j-2}^{j+5}
\alpha_{k}^+ e_{k}^+  = 0
\label{receqeven}
\end{align}
should hold.
Using this, 
$e_{j}^{+}$ with $j \ge 7$ can be written by $e_{0}^+, e_{1}^+, \cdots e_{6}^+$.

\subsection{Test scalar and vector fields}
We can study the test scalar and vector fields similar to the above discussions.
In that cases, the back ground effective potential becomes 
\begin{align}
V^{(0)} = V_s = \frac{\ell (\ell +1)}{r^2} - (1-s^2)\frac{r_H}{r^3},
\end{align}
where $s = 0$ for the test massless scalar fields, $s = 1$ for the test massless vector fields.
Note that $s =2$ corresponds to the Regge-Wheeler potential $V^-$.
We parameterize $\delta V$ as
\begin{align}
\delta V = \delta V_s = \frac{1}{r_H^2} \sum_{j = 0}^\infty \beta_j^s \left(\frac{r_H}{r}\right)^j,
\end{align}
with the small parameters $\beta_s$,
then the QNM frequency becomes
\begin{align}
\omega_{\rm QNM} = \omega_0^s + \sum_{j = 0}^{\infty} \beta_j^s d_j^s,
\end{align}
where the coefficients $d_j^s$ are constants which do not depend on $\beta_j^s$.
Setting
\begin{align}
Y = y_j \left(\frac{r_H}{r}\right)^j,
\end{align}
with an integer $j$ and a constant $y_j$,
after the same calculation as above discussions, we obtain a relation among $d_j^s$ 
\begin{align}
& - 4 j r_H^2 (\omega_0^s)^2 d_{j+1}^s
-
(j+1)(j-2\ell)( j + 2 \ell +2) d_{j+3}^s
\notag\\& +
(2 j + 3)(2 + j(j+3) - 2 \ell (\ell + 1) - 2 s^2)d_{j+4}^s
-
(j+2)(j  + 2 - 2s)(j  +2 + 2s) d_{j+5}^s = 0.
\end{align}
Using this relation, $d_j^s$ with $j \ge 3$ can be written by
$d_0^s, d_2^s$ and $d_7^s$.

\section{reduction of the effective potential to lower oder}
We showed that $e_j^\pm$ with high $j$ can be written by that with low $j$.
In a similar way, we can remove the higher order terms of $r_H/r$ in $\delta V$
by choosing the function $Y$ appropriately.
For simplicity, we only focus on the odd parity case in this section.
Note that the similar discussion holds if we consider the cases of even parity, vector fields or scalar field.

Let us assume that the highest order term in $\delta V$ is 
\begin{align}
 \delta V^{\rm highest} = \frac{\alpha_{n+5}^-}{r_H^2} \left(\frac{r_H}{r}\right)^{n+5}.
\end{align}
If we choose 
\begin{align}
Y = y_n \left(\frac{r_H}{r}\right)^{n},
\end{align}
with
\begin{align}
y_n = \frac{2 r_H \alpha_{n+5}^-}{(n-2)(n+2)(n+6)},
\label{yneq}
\end{align}
we can remove this term, and the new effective potential contains
at most $O((r_H/r)^{n+4})$ terms.
Repeating this process,
we can write the effective potential $\delta V$ in terms of $(r_H/r)^j$ with $j = 0, 1, 2, 7$.\footnote{Note 
that we cannot remove $(r_H/r)^7$ term because Eq.~\eqref{yneq} diverges in that case.
This is similar to why $e_7^-$ cannot be written by $e_0^-$ and $e_2^-$ discussed in footnote.~\ref{foot1}.
}
This implies that we only need to consider the case
\begin{align}
 \delta V = \frac{1}{r_H^2}
\left[
\alpha_0^- 
+
\alpha_1^- \left(\frac{r_H}{r}\right)
+ 
\alpha_2^- \left(\frac{r_H}{r}\right)^{2}
+ 
\alpha_7^- \left(\frac{r_H}{r}\right)^{7}
\right],
\end{align}
instead of Eq.~\eqref{deltav}.
We should note that
the coefficients $\alpha_0^-, \alpha_1^-, \alpha_2^-, \alpha_7^-$
become complicated functions of $\omega_0^-$, in general.

\section{summary and discussion}
In the parametrized black hole quasinormal ringdown formalism~\cite{Cardoso:2019mqo},
the deviation of QNM frequency from GR is given by
$\sum_j \alpha_j^\pm e_j^\pm$.
While the parameters $\alpha_j^\pm$, which depend on physical situations,
can be read from the master equation,
we need to numerically calculate the model independent coefficients $e_j^\pm$ 
in advance~\cite{Cardoso:2019mqo}.
Especially, it was technically not easy task to compute $e_j^\pm$ with high $j$ in a high accuracy.

In this paper, we derived recursion relations among $e_j^\pm$ with different $j$.
Using these relations, $e_j^\pm$ with high $j$ can be written by that with low $j$.
This can efficiently reduce the calculation cost for computing $e_j^\pm$.
The recursion relations can also be used to estimate the numerical calculation error
for $e_j^\pm$. For example, if we numerically calculate $e_0^-, e_2^-, e_3^-$, 
we can estimate the error from $|1 - {\rm LHS}/{\rm RHS}|$, where 
${\rm LHS}$ and ${\rm RHS}$ denote the left and right hand side of Eq.~\eqref{e3m}, respectively.

In a similar way, we also showed that $\delta V$ can be rewritten in terms of 
$(r_H/r)^j$ with only low $j$ using the ambiguity of the effective potential.
It is also interesting to extend the present discussion to the coupled cases and the higher order cases~\cite{McManus:2019ulj}, but we leave them for future work.

\section*{Acknowledgments}
MK would like to thank Yasuyuki Hatsuda for useful discussions and comments on the paper.

\appendix
\section{general discussion of the ambiguity of the effective potential}
\label{generalcase}
Let us consider the master equation with Schr\"odinger form
\begin{align}
\frac{d^2}{dx^2} \Phi + (\omega^2 - V) \Phi = 0,
\label{mastereqV}
\end{align}
where $V$ is the effective potential. 
For the case in~Eq.~\eqref{mastereq}, the relation between $x$ and $r$ is given by $d/dx = f d/dr$.
In this section, we discuss the general differential transformations of the master variable which keep 
the master equation to be Schr\"odinger form.
Defining a new master variable $\chi$ as\footnote{
Because $\Phi$ satisfies the second order differential equation Eq.~\eqref{mastereqV},
the second or more derivatives of $\Phi$ can be written in terms of $\Phi$ and $\Phi^\prime$.
For this reason, we only need to consider differential transformations
which only contain at most first derivative of $\Phi$.
}
\begin{align}
 \chi = A \Phi + B \frac{d\Phi}{dx},
\end{align}
with
\begin{align}
 - B^2 V^\prime + B (2 (\omega^2 - V)B^\prime - A^{\prime \prime}) + A (2 A^\prime + B^{\prime \prime}) = 0,
\label{eqzvy}
\end{align}
$\chi$ satisfies the equation
\begin{align}
\chi^{\prime \prime} + \left(\omega^2 - \left(V + \delta W \right)\right) \chi = 0,
\label{chimastereq}
\end{align}
with
\begin{align}
\delta W = \frac{2 A^\prime}{B} + \frac{B^{\prime \prime}}{B},
\end{align}
where $A, B$ are functions of $x$ and 
$\prime$ denotes derivative with respect to $x$.
Note that Eq.~\eqref{eqzvy} is the condition such that 
the coefficient of $\chi^{\prime}$ vanishes in the master equation.
We can regard the term $\delta W$ as an ambiguity of the effective potential.
Eq.~\eqref{eqzvy} can be integrated as
\begin{align}
 (-\omega^2  + V) B^2 - A^2 + B A^\prime - A B^\prime = c_2,
\label{zyeqc}
\end{align}
with a constant $c_2$.
Defining
\begin{align}
 S := \frac{A}{B} + \frac{B^\prime}{B},
\end{align}
Eq.~\eqref{chimastereq} can be written as 
\begin{align}
\chi^{\prime \prime} + \left( 
-\frac{c_2}{B^2} - S^2 - S^\prime
\right) \chi = 0.
\label{chimastereq2}
\end{align}
\subsection{$A = 1 - \epsilon X$ and $B = -\epsilon Y$ case}
For $A = 1 - \epsilon X$ and $B = - \epsilon Y$ with the functions $X, Y$ and a small parameter $\epsilon$,
because $\chi = \Phi$ at ${\cal O}(\epsilon^0)$, the relation
\begin{align}
\Phi = (1 + \epsilon X)\chi + \epsilon Y \chi^\prime + {\cal O}(\epsilon^2)
\end{align}
holds.
From Eq.~\eqref{zyeqc}, we obtain $X = c_3 - Y^\prime/2 + {\cal O}(\epsilon)$ where $c_3 = (1 - \sqrt{-c_2})/\epsilon$.
Then, $\delta W$ becomes
\begin{align}
\delta W
&=
\frac{B V^\prime - 2 (\omega^2 - V)B^\prime + A^{\prime \prime}}{A}
=
\epsilon\left[
-Y V^\prime + 2 (\omega^2 - V) Y^\prime + \frac{1}{2}Y^{\prime \prime \prime}
\right] + {\cal O}(\epsilon^2),
\end{align}
where we used Eq.~\eqref{eqzvy} in the first equality.
This result is same as Eq.~\eqref{deltaw}.
Note that $d/dx = f d/dr$ and $\delta W$ in this section corresponds to that in Sec.~\ref{secii} multiplied by $f$.

\subsection{$B = 1, c_2 = -\omega^2$ case}
If we set $B = 1$ and $c_2 = -\omega^2$,
Eq.~\eqref{zyeqc} becomes
\begin{align}
V = S^2 - S^\prime
\end{align}
and the master equation Eq.~\eqref{chimastereq2} becomes
\begin{align}
\chi^{\prime \prime} + \left( 
\omega^2 - S^2 - S^\prime
\right) \chi = 0.
\end{align}
This is nothing but the superpartner of the original system Eq.~\eqref{mastereqV}.

\end{document}